\documentclass[ twocolumn,aps,prd,   
               preprintnumbers,numbers,sort&compress,
               nofootinbib,
                            showpacs,
               colorlinks,
               linkcolor=blue,
               citecolor=blue,
               superscriptaddress]{revtex4-1}
   \newcommand{\exclude}[1]{}

\usepackage{graphicx,amsmath,amssymb,bm}
\DeclareMathOperator{\Tr}{Tr}
\usepackage{psfrag}
\usepackage{hyperref}
\usepackage{enumitem}
\usepackage{siunitx}

\usepackage{color}

\newcommand{\beq}{\begin{equation}}
\newcommand{\eeq}{\end{equation}}
\newcommand{\be}{\begin{eqnarray}}
\newcommand{\ee}{\end{eqnarray}}
   
\def\dd{ \,\mathrm{d} }

\def\+{\dagger}
 \def\la{\langle}
 \def\ra{\rangle}

\begin{document}

\title{  Aharonov-Bohm phases 
%and  tunneling transitions 
in a quantum   LC circuit      }

%\author{}

\author{ChunJun Cao}
\affiliation{Walter Burke Institute for Theoretical Physics, California Institute of Technology, Pasadena, CA 91125, USA}
\author{Yuan Yao}
\affiliation{Department of Physics and Astronomy, University of British Columbia, Vancouver, B.C. V6T 1Z1, Canada} 
\author{Ariel R. Zhitnitsky}  
\affiliation{Department of Physics and Astronomy, University of British Columbia, Vancouver, B.C. V6T 1Z1, Canada} 
%\date{\today}

\begin{abstract}
We study novel types of contributions to the partition function of the Maxwell system defined on a small compact manifold.  These contributions, often not addressed in the perturbative treatment with physical photons, emerge as a result of tunneling transitions between topologically distinct but physically identical vacuum winding states. These new terms give an extra contribution to the Casimir pressure, yet to be measured. 
We argue that this effect is highly sensitive to a small external electric field, which should be contrasted with the conventional Casimir effect where the vacuum photons are essentially unaffected by any external field. Furthermore, photons will be emitted from the vacuum in response to a time-dependent electric field, similar to the dynamical Casimir effect in which real particles are radiated from the vacuum due to the time-dependent boundary conditions.
 We also propose an experimental setup using a quantum LC circuit to detect this novel effect. We expect physical electric charges to appear on the capacitor plates when the system dimension is such that coherent Aharonov-Bohm phases can be maintained over macroscopically large distances.

 
\end{abstract} 

\maketitle

\section{Introduction. Motivation.}\label{introduction}
 It has been recently argued \cite{Cao:2013na,Zhitnitsky:2013hba,Zhitnitsky:2014dra,Zhitnitsky:2015fpa} that some novel  terms in the partition function emerge when a pure Maxwell theory is defined on a small compact manifold.  These terms are not related to the  propagating photons with two transverse physical polarizations, which are responsible for the conventional Casimir effect (CE)\cite{Casimir}. Rather, they occur as a result  of  tunneling  events between topologically different but physically identical    $|k\ra$ topological sectors. While such contributions are trivial in Minkowski space-time ${\mathbb{R}_{1,3}}$, they become important when the system is defined on certain small  compact manifolds. Without loss of generality, consider a manifold  ${\mathbb{M}}$
 % which could be the 4-torus ${\mathbb{T}^4}$, 
which has at least one non-trivial direct factor of the fundamental group, e.g., $\pi_1[U(1)]\cong \mathbb{Z}$. The presence of the topological sectors $|k\ra$, which play a key role in our discussions, arise precisely from the presence of such nontrivial mappings for the Maxwell $U(1)$ gauge theory. The corresponding physically observable phenomenon has been termed the topological Casimir effect (TCE). 
  
 In particular,  it has been explicitly shown in  \cite{Cao:2013na}  that these novel terms in  the topological portion of the partition function ${\cal{Z}}_{\rm top}$  lead to a fundamentally new  contribution to the Casimir vacuum pressure that appears as a result  of tunneling events between topological sectors $|k\ra$. 
 Furthermore,   ${\cal{Z}}_{\rm top}$  displays  many features 
 of   topologically ordered systems, which were  initially   introduced in the context of condensed matter systems (see recent reviews \cite{Cho:2010rk,Wen:2012hm,Sachdev:2012dq, Cortijo:2011aa, Volovik:2011kg}):  ${\cal{Z}}_{\rm top}$ demonstrates  the degeneracy of the system which can only be described in terms of non-local operators  \cite{Zhitnitsky:2013hba}; the infrared physics of the system can be studied  in terms of  non-propagating auxiliary topological fields \cite{Zhitnitsky:2014dra}, analogous to how a topologically ordered system  can be  analyzed  in terms of  the Berry's connection (also an emergent rather than fundamental field), and the corresponding expectation value of the  auxiliary topological field  determines the phase of the system. Finally, one can show \cite{Zhitnitsky:2015fpa} that if the same system is placed in a time-dependent magnetic field $B_{\rm ext}^z(t)$,  real photons will be emitted from the vacuum, similar to the so-called dynamical Casimir effect (DCE) \cite{DCE, DCE-review, DCE-exp}).  The difference from the conventional DCE is that the dynamics of the vacuum in our system defined on a small compact manifold is not related to the fluctuations of the conventional degrees of freedom, the virtual photons. Rather, the radiation here arises from tunneling events between topologically different but physically identical $|k\ra$  sectors in a time-dependent background. 
 
 As we review in section \ref{magnetic},  the relevant   vacuum fluctuations  which saturate the topological portion of the partition function ${\cal{Z}}_{\rm top}$ are formulated in terms of topologically nontrivial boundary conditions. Classical instantons formulated in Euclidean space-time satisfy the periodic boundary conditions up to a large gauge transformation and  provide a topological magnetic instanton-flux in the $z$-direction.
 These integer magnetic fluxes describe the tunneling transitions between physically identical but topologically distinct $|k\ra$ sectors. 
 Precisely these field configurations generate   an  extra Casimir vacuum pressure in the system. What happens to this complicated vacuum structure when the system is placed in the background of a constant external magnetic field  $B_{\rm ext}^z$? The answer is known \cite{Cao:2013na}: the corresponding partition function ${\cal{Z}}_{\rm top}$ as well as all  observables, including the topological contribution to the Casimir pressure,  are highly sensitive to small magnetic fields and demonstrate $2\pi$ periodicity  with respect to magnetic flux represented by the parameter $\theta_{\rm eff}\equiv eSB_{\rm ext}^z$  where $S$ is the $xy$ area   of the system  ${\mathbb{M}}$.  This sensitivity to external magnetic field  is a result of  the quantum interference  of the external field with the topological quantum fluctuations. Alternatively, one can see this as resulting from a small but non-trivial overlap between the conventional Fock states, constructed by perturbative expansions around each $|k\rangle$ sector, and the true energy eigenstates of the theory, which are only attainable in a non-perturbative computation that takes the tunneling into account. This strong ``quantum" sensitivity of the TCE should be contrasted  with conventional Casimir forces  which are  practically unaltered by any external    field due to the strong suppression $\sim B_{\rm ext}^2/m_e^4$ (see   \cite{Cao:2013na} for the details).

The main goal of the present work is  to study  the dynamics of electric instanton-fluxes in contrast with   the magnetic ones considered previously in \cite{Cao:2013na,Zhitnitsky:2013hba,Zhitnitsky:2014dra,Zhitnitsky:2015fpa}. Placing metallic plates at the opposing ends in the $z$-direction endows our system with the geometry of a small quantum capacitor. Periodic boundary conditions up to a large gauge transformation can be enforced by connecting the two plates of the capacitor with an external wire. Thus, the system forms a quantum LC circuit where the wire provides the inductance. Formal computation of the partition function with electric-type instantons is very similar to previous studies. Like their magnetic counterparts, these electric instantons also describe tunneling transitions between physically identical but topologically distinct $|k\ra$ sectors in Euclidean space-time. Both types of instanton-fluxes give rise to an extra repulsive pressure, the opposite of the conventional CE, and both exhibit oscillatory behavior in pressure, induced field, and susceptibility in response to an external field represented by an effective theta parameter, $\theta_{\rm eff}$.

 However, the behaviour of the system as well as the physical interpretation drastically changes when one considers the system placed in an external field: while the magnetic field in  space-time with  Euclidean or  Minkowski  signatures remains unaltered, the electric fluxes pick up an imaginary $i$ in the transition from Minkowski to Euclidean space-time. Most notably, whereas the magnetic oscillations have universal $2 \pi$-periodicity for all system sizes, the periodicity in the electric system varies with the system size. Compared to the magnetic system, an advantage of the electric system is that the system size can be much more easily optimized to produce topological effects of order one.

The topological instanton configurations which describe the tuneling transitions are formulated in terms of periodic boundary conditions on the gauge field up to a large gauge transformation. These boundary conditions correspond to persistent fluctuating charges that reside on the capacitor plates. These charges can be interpreted as a consequence of the Aharonov-Bohm phases of the topological vacuum. Positive and negative charges on each plate  cancel exactly in the absence of an external electric field, but with a non-vanishing external field, each plate acquires a nonzero net charge, giving rise to an induced dipole moment. 

We also study the emission of real photons from the topological vacuum when the external electric field varies slowly (in the adiabatic limit). The external field interferes with the topological vacuum configurations and generates real photons whose energy ultimately comes from the time-varying external field. One can also explain this emission in terms of the dipole moment generated by the fluctuating electric charges. As the external field takes on time-dependence, so does the induced dipole moment. And a time-varying dipole moment naturally implies electromagnetic radiation.

The structure of our presentation is as follows. In section \ref{magnetic}, we review the construction of magnetic-type instantons and the partition function $\cal{Z}_{\rm top}$ in \cite{Cao:2013na,Zhitnitsky:2013hba,Zhitnitsky:2014dra,Zhitnitsky:2015fpa}. We also explain how an external magnetic field enters $\cal{Z}_{\rm top}$. In section \ref{electric}, we construct the electric-type instantons that are the focus of the present studies, and their extra vacuum energy contribution. In section \ref{external field}, we consider the system placed in an external electric field. In particular, we explain in detail how to correctly treat an external field applied in Minkowski space-time in the Euclidean description. We also compute the various physical parameters of the system such as the induced electric field and corresponding  susceptibility which measures the response of the system with respect to the applied field. 
In section \ref{dipole}, we provide an alternative interpretation of the topological vacuum configurations in terms of fluctuating boundary charges, and comment on the connection with a previously discussed phenomenon  of  persistent currents. In 
 section \ref{numerics} we propose an experimental setup using quantum LC circuits to detect these topological effects, and numerically estimate the magnitude of the boundary charges and the associated dipole moment in systems with experimentally accessible dimensions.  
In section \ref{radiation}   we discuss the emission of real photons as the static external field takes on time-dependence. And in section \ref{conclusion}, we conclude the present studies and speculate on the possible relevance of the TCE for cosmology. In particular, the de Sitter behavior in the inflationary epoch could be an inherent property of the topological sectors in QCD in the expanding universe, rather than a result of some ad hoc dynamical field such as inflaton. The emission of real physical degrees of freedom from the inflationary vacuum in a time-dependent background (the so-called reheating epoch) in all respects resembles the effect considered in the present work where real photons can be emitted from the vacuum in a time-varying external electric field.

  \section{Topological partition function. Magnetic type instantons}\label{magnetic}
 Our goal here is to review the Maxwell system  defined  on a Euclidean 4-torus   with  sizes $L_1 \times L_2 \times L_3 \times \beta$ in the respective directions. This 4-torus provides the infrared regularization of the system, which plays a key role in the proper treatment of the  topological terms  related to tunneling events between topologically distinct but physically identical $|k\ra$ sectors.

 We follow  \cite{Cao:2013na}  in our construction of the partition function ${\cal{Z}}_{\rm top}$ where it was employed to compute the corrections to the Casimir effect due to these topological fluctuations. The crucial point is that we impose periodic boundary conditions on the gauge field $A^{\mu}$ up to a large gauge transformation.
 In what follows we simplify our analysis by considering   a clear case with winding topological sectors $|k\ra$    in the $z$-direction only.  The classical instanton configuration in Euclidean space  which describes the corresponding tunneling transitions can be represented as follows:
\be
\label{toppot4d}
A^{\mu}_{\rm top} = \left(0 ,~ -\frac{\pi k}{e L_{1} L_{2}} x_2 ,~ \frac{\pi k}{e L_{1} L_{2}} x_1 ,~ 0 \right),
\ee  
where $k$ is the winding number that labels the topological sector, and $L_{1}$, $L_{2}$ are the dimensions of the plates in the $x$- and $y$-directions respectively, assumed to be much larger than the plate separation in the $z$-direction, $L_3$. The terminology ``instanton" is adapted   from  similar 
 studies in 2d QED    \cite{Cao:2013na} where the corresponding configuration in the $A_0=0$ gauge describes the interpolation between pure gauge vacuum winding states $|k\ra$. We  use the same terminology and interpretation for the 4d case  because (\ref{topB4d}) is the classical configuration saturating the partition function ${\cal{Z}}_{\rm top}$, in close analogy with the 2d case (details in \cite{Cao:2013na}). This classical instanton  configuration satisfies the periodic boundary conditions up to a large gauge transformation,  and provides a topological magnetic instanton-flux in the $z$-direction:
\be
\label{topB4d}
\vec{B}_{\rm top} &=& \vec{\nabla} \times \vec{A}_{\rm top} = \left(0 ,~ 0,~ \frac{2 \pi k}{e L_{1} L_{2}} \right),\\
\Phi&=&e\int dx_1dx_2  {B}_{\rm top}^z={2\pi}k. \nonumber
\ee
The Euclidean action of the system is quadratic and has the  form  
\be
\label{action4d}
\frac{1}{2} \int \dd^4 x \left\{  \vec{E}^2 +  \left(\vec{B} + \vec{B}_{\rm top}\right)^2 \right\} ,
\ee
where $\vec{E}$ and $\vec{B}$ are the dynamical quantum fluctuations of the gauge field.  
We call the configuration given by Eq. (\ref{toppot4d}) the instanton-fluxes  describing the tunneling events between topological sectors $|k\ra$. These configurations saturate the partition function (see (\ref{Z4d}) below) and should be interpreted as ``large" quantum fluctuations which change the winding states $|k\ra$, in contrast with ``small" quantum fluctuations which are topologically trivial and are expressed in terms of conventional virtual photons saturating  the quantum portion of the partition function ${\cal{Z}}_{\rm quant}$.

The key point is that the  topological portion ${\cal{Z}}_{\rm top}$   decouples from the quantum fluctuations,  ${\cal{Z}} = {\cal{Z}}_{\rm quant} \times {\cal{Z}}_{\rm top}$, such that the quantum fluctuations do not depend on topological sectors $|k\ra$ and can be computed in the trivial topological sector, $k=0$.
Indeed,  the cross term vanishes, 
\be
\int \dd^4 x~ \vec{B} \cdot \vec{B}_{\rm top} = \frac{2 \pi k}{e L_{1} L_{2}} \int \dd^4 x~ B_{z} = 0, 
\label{decouple}
\ee
because the magnetic portion of the quantum fluctuations in the $z$-direction, represented by $B_{z} = \partial_{x} A_{y}  - \partial_{y} A_{x} $, is a periodic function as   $\vec{A} $ is periodic over the domain of integration. 
This technical remark in fact greatly simplifies our  analysis as the contribution of the physical propagating photons 
is not sensitive to the topological sectors. This is,  of course,  a specific feature  of quadratic action 
 (\ref{action4d}), in contrast with non-abelian  and non-linear gauge field theories where quantum fluctuations do depend on the topological sectors.  
  
The classical action  for configuration (\ref{topB4d}) then takes the form 
\be
\label{action4d2}
\frac{1}{2}\int \dd^4 x \vec{B}_{\rm top}^2= \frac{2\pi^2 k^2 \beta L_3}{e^2 L_1 L_2}.
\ee
To further simplify our analysis in  computing  ${\cal{Z}}_{\rm top}$, we consider a geometry where $L_1, L_2 \gg L_3 , \beta$, similar to the construction of the conventional CE.   
In this case our system   is closely related to 2d Maxwell theory by dimensional reduction: taking a slice of the 4d system in the $xy$-plane will yield precisely the topological features of the 2d torus considered in  great detail in  \cite{Cao:2013na}.
  Furthermore, with this geometry our simplification (\ref{topB4d}) where we consider exclusively the magnetic instanton-fluxes in the $z$-direction is justified as the corresponding classical action (\ref{action4d2}) assumes a minimum  possible value.  With this assumption we can consider very low temperatures, but still we cannot take the formal limit $\beta\rightarrow\infty$  in the final expressions because of the technical constraints. 
      
With these additional simplifications   the topological partition function becomes
\be
\label{Z4d}
{\cal{Z}}_{\rm top} = \sqrt{\frac{2\pi \beta L_3}{e^2 L_1 L_2}} \sum_{k\in \mathbb{Z}} e^{-\frac{2\pi^2 k^2 \beta L_3}{e^2 L_1 L_2} }= \sqrt{\pi \tau} \sum_{k\in \mathbb{Z}}e^{-\pi^2 \tau k^2}, ~~~~
\ee
where we have introduced the dimensionless system size parameter
\be
\label{tau}
\tau \equiv {2 \beta L_3}/{e^2 L_1 L_2}.
\ee
Eq. (\ref{Z4d}) is essentially the dimensionally reduced expression of the topological partition function  for the 2d 
 Maxwell theory analyzed in \cite{Cao:2013na}. 
   One should also note that the normalization factor $\sqrt{\pi \tau}$ which appears in (\ref{Z4d}) does not depend on the topological sectors $|k\ra$, and essentially represents our  normalization convention  ${\cal{Z}}_{\rm top}
\rightarrow 1$ in the limit $L_1L_2\rightarrow \infty$, which corresponds to  a convenient setup for  Casimir-type experiments. The simplest way to  demonstrate that ${\cal{Z}}_{\rm top} \rightarrow 1$ in the limit $\tau\rightarrow 0$ is to use the  dual representation  (\ref{Z_dual1}), see below.

Next, we introduce an external magnetic field to the Euclidean Maxwell system. Normally, in the conventional quantization of electromagnetic fields in infinite Minkowski space, there is no \emph{direct} coupling between the fluctuating vacuum photons and an external magnetic field due to the linearity of the Maxwell equations. Coupling with  fermions generates  a negligible effect $\sim \alpha^2B_{\rm ext}^2/m_e^4$ as the non-linear Euler-Heisenberg effective Lagrangian  suggests (see \cite{Cao:2013na} for the details and numerical estimates).  In contrast, the external magnetic field does couple with the topological fluctuations (\ref{topB4d}) and can lead to effects of order unity.

The corresponding partition function can be easily constructed for the external magnetic field $B^{z}_{\rm ext}$ pointing along the $z$-direction, as 
the crucial decoupling of the background field  from the quantum fluctuations assumes the same form (\ref{decouple}).  In other words, the physical propagating photons with non-vanishing momenta are not sensitive to the topological sectors $|k\ra$, nor to the external magnetic field, similar to the discussions after Eq. (\ref{decouple}). Additionally, since a real-valued external magnetic field applied in Minkowski space-time remains the same after analytic continuation to Euclidean space-time, this $B^{z}_{\rm ext}$ can be used to represent both the Minkowski external field and the Euclidean one.

The classical action in the presence of this uniform static magnetic field $B^{z}_{\rm ext}$ therefore takes the form 
\be
\label{B_ext}
\frac{1}{2}\int \dd^4 x  \left(\vec{B}_{\rm ext} + \vec{B}_{\rm top}\right)^2=  \pi^2\tau\left(k+\frac{\theta_{\rm eff}}{2\pi} \right)^2
\ee
where the effective theta parameter $\theta_{\rm eff} \equiv e L_1L_2 B^z_{\rm ext}$ is defined in terms of the external magnetic field $B^z_{\rm ext}$.
And the partition function can be easily reconstructed from (\ref{Z4d}):
\be 
\label{Z_eff}
  {\cal{Z}}_{\rm top}(\tau, \theta_{\rm eff})
 =\sqrt{\pi\tau} \sum_{k \in \mathbb{Z}} \exp\left[-\pi^2\tau \left(k+\frac{\theta_{\rm eff}}{2\pi}\right)^2\right].~~
\ee
 The dual representation for the partition function is obtained by applying the Poisson summation formula  such that (\ref{Z_eff}) becomes 
  \be 
\label{Z_dual1}
  {\cal{Z}}_{\rm top}(\tau, \theta_{\rm eff})
  = \sum_{n\in \mathbb{Z}} \exp\left[-\frac{n^2}{\tau}+in\cdot\theta_{\rm eff}\right]. 
  \ee
 Eq. (\ref{Z_dual1})  justifies our notation for  the effective theta parameter $\theta_{\rm eff}$ as it enters the partition function in combination with integer $n$. One should emphasize that the $n$ in the dual representation (\ref{Z_dual1}) is not the integer magnetic flux $k$ defined in Eq. 
(\ref{topB4d}) which enters the original partition function (\ref{Z4d}). Furthermore,  the $\theta_{\rm eff}$ parameter which enters (\ref{Z_eff}, \ref{Z_dual1}) is not the fundamental $\theta$ parameter normally introduced into the Lagrangian  in front of the $\vec{E}\cdot\vec{B}$ operator. Rather,  $\theta_{\rm eff}$ should be understood as an effective parameter representing the construction of the  $|\theta_{\rm eff}\ra$ state for each slice with non-trivial $\pi_1[U(1)]$ in the 4d system. In fact, there are three such  $\theta_{\rm eff}^{M_i}$  parameters representing different slices of the 4-torus and their corresponding external magnetic fluxes. There are similarly three $\theta_{\rm eff}^{E_i}$
parameters representing the external electric fluxes (in Euclidean space-time)   as discussed  in \cite{Zhitnitsky:2013hba} and as we will discuss in section \ref{external field}, such that the total number of $\theta$ parameters classifying the system is six, in agreement with the total number of hyperplanes in four dimensions\footnote{Since it is not possible to have a 3D spatial torus without embedding it in 4D spatial space, the corresponding construction where all six possible types of  fluxes are generated represents a pure academic interest.\label{torus}}. 

We shall not elaborate on this classification in the present work. Instead,  we limit ourselves to a single $\theta_{\rm eff}^{E_z}$ describing the partition function in the presence of an (Euclidean) external electric field $ E^z_{\rm ext} $ pointing  in the $z$-direction. So essentially in what follows, we consider the manifold $\mathbb{I}^1\times \mathbb{I}^1\times \mathbb{S}^1\times \mathbb{S}^1$, where $\mathbb{I}^1$ is an    interval in the $x$- and $y$-directions with length $L_1$ and $L_2$ respectively, while  a single  spatial circle $\mathbb{S}^1$ with size $L_3$ points in the $z$-direction.  Later in the paper, we interpret the obtained results by switching back to the physical  Minkowski space-time. As we shall see, there are significant differences in the behavior of the system in comparison with the previously considered magnetic case  \cite{Cao:2013na}.

\section{Electric type instantons} \label{electric}
We formulate the electric system on a  Euclidean 4-manifold $\mathbb{I}^1\times \mathbb{I}^1\times \mathbb{S}^1\times \mathbb{S}^1$  with size $L_1\times L_2 \times L_3\times \beta$. \exclude{Some of the frequent questions I got from experimentalists are how to make a 4 torus. I understand that we wish to keep it general at the beginning, but we don't actually talk about the other 5 terms for a config like 4-torus. So I guess if we wish to get their interest, it will make more sense to just start with simpler Euclidean topology for experiments, like $S^1\times S^1\times I^1\times I^1$, where $I^1$ is an  interval. Namely we have one spatial circle. In this case the topological partition function for this one term remains the same and actually is exact for this system. We also have clean decoupling but now we can actually give physical setups like a hollow annulus with rectangular cross section. EM modes in this case is definitely periodic in z, and have some BC for x and y direction. This now looks like a Casimir experiment with different shapes. The difference is we ask them to measure the electric response when external field is applied. Granted that conventional modes change a bit, but won't be a problem since we don't discuss them here.} Two parallel conducting plates form the boundary in the $z$-direction, endowing the system with the geometry of a small quantum capacitor that has plate area $L_1 \times L_2$ and separation $L_3$ at an ambient temperature of $T=1/\beta$. \exclude{This parallel plate description still confuses me, and chances are, it will confuse the readers as well..I mean we have periodicity in the z-direction, the introduction of plates/capacitor is quite misleading in the way that it is not strictly a capacitor...it is two plates with a wire connecting them to enforce the PBC. If we use the simpler setup proposed above, then I think it is more clear to remove the capacitor description. Similar arguments apply to the title... In this section we do not discuss the question  on possible practical realizations of relevant geometry when  nontrivial mapping  $\pi_1[U(1)]$ with electric fluxes may occur.  Similar question on    nontrivial magnetic  mapping $\pi_1[U(1)]$ was discussed   in ref. \cite{Zhitnitsky:2015fpa} where it was  argued that  a simple  geometry in form of a ring is sufficient to  generate   nontrivial topological magnetic fluxes along $z$ direction with a single classification parameter $\theta_{\rm eff}$. We postpone the  corresponding discussions for the electric case until section \ref{numerics}. } These two plates are connected by an external  wire to enforce the periodic boundary conditions (up to large gauge transformations) in the $z$-direction, and so the system can be viewed as a quantum LC circuit where the external wire forms an inductor L. The quantum vacuum  between the plates (where the tunneling transitions occur) represents  the  object of our studies.

\subsection{Construction of topological partition function}

The classical instanton configuration in Euclidean space-time which describes tunneling transitions between the topological sectors $|k\ra$ can be represented as follows:
\be
\label{A_top}
A^{\mu}_{\rm top} (t) &=& \left(0 ,~ 0,~ 0, ~ \frac{2\pi k}{e L_{3} \beta } t  \right) \\  \nonumber
A^3_{\rm top}(\beta)&=&A^3_{\rm top}(0)+\frac{2\pi k}{eL_3},
\ee  
where $k$ is the winding number that labels the topological sector and $t$ is the Euclidean time.  This classical instanton  configuration satisfies the periodic boundary conditions up to a large gauge transformation,  and produces a topological electric instanton-flux in the $z$-direction:
\be
\label{E_top}
\vec{E}_{\rm top} =\dot{\vec{A}}_{\rm top} = \left(0 ,~ 0,~ \frac{2 \pi k}{e L_{3} \beta} \right).
 \ee
 This  construction of these electric-type instantons is in fact much closer (in  comparison with the magnetic instantons  discussed  in \cite{Cao:2013na}) to the  Schwinger model on a circle
 where the relevant instanton configurations were originally constructed \cite{SW}. 
 The Euclidean action of the system takes the form 
\be
\label{action_E}
\frac{1}{2} \int \dd^4 x \left\{  \left(\vec{E} + \vec{E}_{\rm top} \right)^2  +  \vec{B}^2 \right\} ,
\ee
where, as in the magnetic case, $\vec{E}$ and $\vec{B}$ are the dynamical quantum fluctuations of the gauge field. Because periodic boundary conditions have been imposed on the system, the topological and quantum portions of the partition function again decouple: ${\cal{Z}} = {\cal{Z}}_{\rm quant} \times {\cal{Z}}_{\rm top}$. One can explicitly check that the cross term vanishes:
\be
\int \dd^4 x~ \vec{E} \cdot \vec{E}_{\rm top} = \frac{2 \pi k}{e \beta L_{3}} \int \dd^4 x~ E_{z} = 0, 
\ee
since $E_z=\partial_0 A_z -\partial_z A_0$ and $\vec{A}$ is periodic over the domain of integration. Hence, the classical action for configuration (\ref{A_top}) becomes
\be
\label{action_E}
\frac{1}{2}\int \dd^4 x \vec{E}_{\rm top}^2= \frac{2\pi^2 k^2 L_1 L_2}{e^2 L_3 \beta}=\pi^2 k^2 \eta
\ee
where $\eta$ is the key parameter characterizing the size of this electric system, defined as 
\be
\label{eta}
\eta\equiv \frac{2L_1L_2}{e^2\beta L_3}. 
\ee
This dimensionless parameter  is related to the $\tau$ parameter in the magnetic case  (\ref{tau}) by  $
\eta =4/e^4\tau$.
With topological action (\ref{action_E}), we next follow the same procedure as in the magnetic case  to construct the topological partition function,
\be
\label{Z_E}
  {\cal{Z}}_{\rm top} (\eta)
 =  \sum_{k \in \mathbb{Z}} e^{-\pi^2\eta k^2},~~
\ee
with normalization ${\cal{Z}}_{\rm top}(\eta\rightarrow\infty)=1$, such that no topological effect survives in the limit $L_1L_2 \rightarrow \infty$.  As a result, Eq.(\ref{Z_E}) differs from the partition fuction in the magnetic case (\ref{Z_E}) by a $k$-independent prefactor. Since the total partition function is represented by  the direct product,  ${\cal{Z}} = {\cal{Z}}_{\rm quant} \times {\cal{Z}}_{\rm top}$, any  $k$-independent factor in the normalization of ${\cal{Z}}_{\rm top}$ can be moved to  ${\cal{Z}}_{\rm quant}$.

The Poisson summation formula can be invoked to obtain the dual expression for the partition function:
\be
\label{Z_E_dual}
  {\cal{Z}}_{\rm top}(\eta)
 =\frac{1}{\sqrt{\pi\eta}}\sum_{n\in \mathbb{Z}} e^{-\frac{n^2}{\eta}}.~~
\ee

\subsection{Topological Casimir pressure}\label{pressure}

\begin{figure}
\centering
\includegraphics[width=0.47\textwidth]{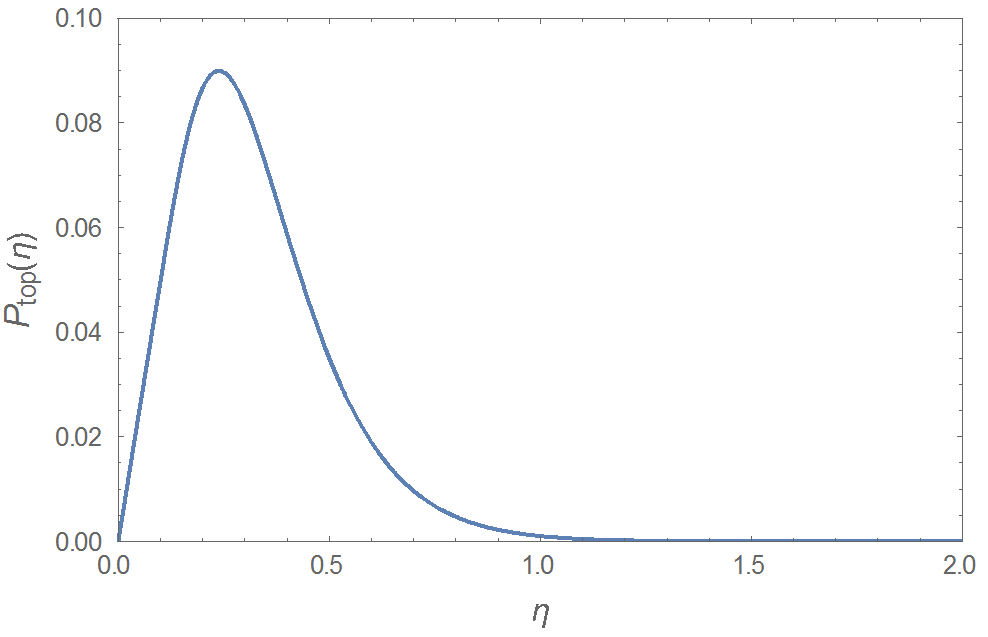}
\caption{\label{fig:pressure_0}A numerical plot of the topological Casimir pressure as a function of the system size parameter $\eta=2 L_1 L_2/e^2 \beta L_3$. Maximum pressure is observed at $\eta \approx 0.25$. Here the pressure is shown in units of $e^2/2L_1^2 L_2^2$.}
\end{figure}

The topological pressure between the capacitor plates can be obtained by differentiating the free energy of the system with respect to plate separation. First, we consider the pressure in the absence of an external electric field:
\be 
\label{P_0}
&\,& P_{\rm top} (\eta)= 
\frac{1}{\beta L_1 L_2} \frac{\partial}{\partial L_3} \ln  {{\cal{Z}}}_{\rm top}(\eta)\\
&=& \frac{e^2}{2L_1^2 L_2^2} \frac{\pi^2 \eta^2}{{ {\cal{Z}}}_{\rm top}(\eta)} \sum_{k \in \mathbb{Z}} k^2 e^{-\pi^2\eta k^2}, \nonumber
\ee
plotted in Fig. \ref{fig:pressure_0} as a function of the system size parameter $\eta$. The pressure peaks around $\eta \approx 0.25$ and has a narrower window compared to the magnetic case  \cite{Cao:2013na}. And like in the magnetic case, this pressure is repulsive, the opposite of the attractive conventional Casimir pressure. In the large $\eta$ limit one can keep only two terms in expression (\ref{P_0}) with $k=\pm 1$, and the pressure reduces to 
\be 
\label{P_0 asymp}
  P_{\rm top}
\approx \frac{e^2 \pi^2 \eta^2}{L_1^2 L_2^2}  e^{-\pi^2\eta},
\ee
with the familiar exponential suppression $\exp[-1/e^2]$ representing the typical behavior of tunneling processes.

To get a sense of the magnitude of this topological pressure, we compare it to the well-known expression for the conventional Casimir pressure between two parallel conducting plates with separation $L_3$:
\be 
  P=- \frac{\pi^2}{240 L_3^4}.
\ee
From Fig. \ref{fig:pressure_0}, the maximum topological pressure (corresponding to $\eta \approx 0.25$) is about $P_{\rm top}^{\rm max} \approx 0.1 e^2/2 L_1^2 L_2^2$, so the maximum ratio between the two pressures is 
\be 
\label{ratio}
  R_{\rm max}=\frac{|P_{\rm top}^{\rm max}|}{|P|} \approx \frac{12 e^2 L_3^4}{\pi^2 L_1^2 L_2^2}=\frac{48 \alpha}{\pi} \frac{L_3^4}{L_1^2 L_2^2}.
\ee
This ratio, even at its maximum, is very small in a typical Casimir experiment setup where $L_1,L_2 \gg L_3$; besides, the numerical prefactor further suppresses it by an order of magnitude.

To sum up this section, we have found an additional contribution to the Casimir pressure that cannot be attributed to any physical propagating degrees of freedom but instead results from the topological excitation of the gauge field. Specifically, this contribution occurs when the system is defined on a compact manifold with nontrivial boundary conditions. As these topological tunneling transitions are described in terms of integer electric fluxes in Euclidean space-time (\ref{E_top}), they exhibit exponential suppression in the conventional geometry, $\eta\rightarrow\infty$ (\ref{P_0 asymp}). And even at its maximum, this topological pressure is orders of magnitude smaller than the conventional Casimir pressure (\ref{ratio}). However, we show in the next section that a unique feature of the topological effect is that an external electric field effectively couples with the electric instanton-fluxes (\ref{E_top}), making the topological effect highly sensitive to an applied electric field. Similar sensitivity to external fields is absent from the conventional CE, where the linearity of the Maxwell equations forbids vacuum photon fluctuations from coupling to external fields.

\section{\texorpdfstring{$\theta$}{TEXT}  vacua and external electric fields} \label{external field}

As shown in the previous section, the pressure produced by the topological CE is many orders of magnitude smaller than the conventional Casimir pressure, making its measurement very difficult. We know that in the magnetic case, as reviewed in section \ref{magnetic}, a weak external magnetic field interferes with the integer magnetic fluxes describing the tunneling events (\ref{topB4d}) and  enters the partition function (\ref{Z_eff}, \ref{Z_dual1}) as an effective theta parameter, eventually producing oscillatory behavior in the physical observables \cite{Cao:2013na}. In this section, we consider the effect of similarly placing the quantum capacitor in a uniform external electric field in the $z$-direction. However, whereas a magnetic field stays unchanged under analytic continuation between Minkowski and Euclidean formulations, an electric field acquires an additional factor of $i$ as it involves the zeroth component of four-vectors, i.e. $E_z=\partial_0 A_z - \partial_z A_0$.

First we consider an external electric field in Euclidean space-time, which simply adds an $\vec{E}_{\rm ext}$ term to the topological action (\ref{action_E}). Note that the quantum fluctuations still decouple from the topological and external fields due to the periodicity of the former over the domain of integration:
\be
\label{decouple2}
\left(E_{\rm ext}^{z} + \frac{2 \pi k}{e \beta L_3} \right) \int \dd^4 x E_{z} = 0.
\ee
The partition function then becomes   
\be 
\label{Z_E_theta}
  {\cal{Z}}_{\rm top}(\eta, \theta_{\rm eff}^E)
 =  \sum_{k \in \mathbb{Z}} \exp\left[-\pi^2\eta \left(k+\frac{\theta_{\rm eff}^E}{2\pi}\right)^2~\right],~~
\ee
where the external Euclidean electric field enters the partition function  through the combination
\be
\label{theta}
\theta_{\rm eff}^E=eL_3\beta E_{\rm ext}.
\ee

In what follows we also need a normalization at non-vanishing external field. Since the portion of the partition function proportional to $E_{\rm ext}^2$ is also $k$-independent, we move it to ${\cal{Z}}_{\rm quant}$. To avoid confusion with notation we use  
$\bar{{\cal{Z}}}_{\rm top}(\eta, \theta_{\rm eff}^E)$ for the partition function with this term removed. It is likewise normalized to one in the large $\eta$ limit in the background of a non-vanishing external source, i.e.

\be
\label{normalization}
 &&{\cal{Z}}_{\rm top}(\eta, \theta_{\rm eff}^E)\equiv \exp\left[- \frac{\eta(\theta_{\rm eff}^E)^2}{4} \right] \times \bar{{\cal{Z}}}_{\rm top}(\eta, \theta_{\rm eff}^E)\nonumber\\
&&\bar{{\cal{Z}}}_{\rm top}(\eta, \theta_{\rm eff}^E) \equiv \sum_{k \in \mathbb{Z}} \exp\left[-\pi^2\eta \left(k^2+\frac{k~\theta_{\rm eff}^E}{\pi}\right)~\right] \nonumber\\
&&\bar{{\cal{Z}}}_{\rm top}(\eta\rightarrow\infty, \theta_{\rm eff}^E) =1.
\ee
One can interpret $\bar{{\cal{Z}}}_{\rm top}(\eta, \theta_{\rm eff}^E)$ as the partition function with the external source contribution $\frac{1}{2}E_{\rm ext}^2 \beta V= \frac{1}{4}\eta(\theta_{\rm eff}^E)^2$ removed from the free energy of the system (see Appendix. \ref{Hamiltonian}  for further justification in the Hamiltonian formulation).  Our normalization $\bar{{\cal{Z}}}_{\rm top}(\eta\rightarrow\infty, \theta_{\rm eff}^E) =1$ corresponds to the geometry when tunneling events are strongly suppressed, i.e., physical phenomena discussed in the present work are trivial for systems in such limit. 

The dual representation for the partition function is obtained by applying the Poisson summation formula  such that (\ref{Z_E_theta}), (\ref{normalization})  become 
  \be 
\label{Z_E_dual}
  {\cal{Z}}_{\rm top}(\eta, \theta_{\rm eff}^E)
  &=& \frac{1}{\sqrt{\pi\eta}}\sum_{n\in \mathbb{Z}} \exp\left[-\frac{n^2}{\eta}+in\cdot\theta_{\rm eff}^E\right] \nonumber\\
  \bar{{\cal{Z}}}_{\rm top}(\eta, \theta_{\rm eff}^E) &=&  \exp\left[\frac{\eta(\theta_{\rm eff}^E)^2}{4} \right] \times{\cal{Z}}_{\rm top}(\eta, \theta_{\rm eff}^E).
  \ee

Unfortunately, we cannot directly calculate physically meaningful thermodynamic properties of the system from this partition function, since ${\theta_{\rm eff}^E}$ does not represent a physical electric field living in Minkowski space-time. Rather, we need to first switch to a Minkowski field by the formal replacement $E_{\rm Euclidean}\rightarrow i E_{\rm Minkowski}$. Explicitly, the partition function in the presence of a real Minkowski electric field is given by   
\be 
\label{Z_M}
  \bar{{\cal{Z}}}_{\rm top}(\eta, \theta_{\rm eff}^M)
 =   \sum_{k \in \mathbb{Z}} \exp{\left[-\eta\left(\pi^2 k^2+i  \pi k \theta_{\rm eff}^M   \right)\right]},~~
\ee
where the physical Minkowski electric field $ E^{\rm Mink}_{\rm ext}$ enters the partition function $ {\cal{Z}}_{\rm top}(\eta, \theta_{\rm eff}^M)$  through the combination
\be
\label{theta_M}
\theta_{\rm eff}^M=eL_3\beta E^{\rm Mink}_{\rm ext}=-i\theta_{\rm eff}^E.
\ee
Our interpretation in this case remains the same: in the presence of a physical external electric field $E^{\rm Mink}_{\rm ext}$ represented by the complex source ${\theta_{\rm eff}^E}$, the path integral (\ref{Z_E_theta}) is saturated by the Euclidean configurations (\ref{E_top}) describing physical tunneling events between the topological sectors $|k\ra$.

With an external Minkowski electric field, the $z$-direction topological pressure computed from the partition function (\ref{Z_M}) exhibits oscillatory behaviour with respect to $\theta_{\rm eff}^M$.  A notable feature of the pressure is that the oscillation period  depends  on the dimensionless parameter $\eta$ characterizing the system (\ref{eta}), in marked contrast with the magnetic case \cite{Cao:2013na} where there is a universal $2 \pi$-periodicity for all values of $\tau$.  

To illustrate the $ \theta_{\rm eff}^M$-dependence  of the observables, first consider a simple case with  $\eta \gg 1$, where the magnitude of all effects are exponentially small.  In this limiting case for sufficiently small external field $E^{\rm Mink}_{\rm ext}$ one can keep only the lowest branch with $k=\pm 1$ in Eq. (\ref{Z_M}) 
  such that the expression for the pressure assumes the form
 \be
 \label{pressure_theta}
 &&P_{\rm top} (0 \leq \eta \pi \theta_{\rm eff}^M\leq \pi)
\approx \frac{e^2 \pi^2 \eta^2}{L_1^2 L_2^2} \exp{(-\pi^2\eta)}\\
&& \times\left[ \cos(\pi \eta \theta_{\rm eff}^M) +\frac{(\pi\eta\theta_{\rm eff}^M)}{\pi^2\eta} \sin(\pi \eta \theta_{\rm eff}^M) \right]. ~~ \nonumber 
\ee
 This formula is   valid only for the first branch when the electric field is small,  $ \eta \pi |\theta_{\rm eff}^M|\leq \pi$. When the external field becomes stronger 
one should be more careful with formal differentiation of the partition function  (\ref{Z_M}) with respect to external parameters. This is because the partition function  is a   periodic function  of    $\theta_{\rm eff}^M$, and this periodicity must be respected by all physical observables\footnote{\label{degeneracy}To be more precise, the physics is periodic under the shift  $\pi \eta\theta_{\rm eff}^M \rightarrow \pi \eta\theta_{\rm eff}^M + 2\pi m$.  Similar  periodicity with respect to an external parameter is quite generic in many gauge theories, including  QCD,  where all physical observables are  periodic functions of the fundamental $\theta$ parameter. This periodicity can be formally enforced in our case  in the path integral by replacing  $\pi \eta\theta_{\rm eff}^M \rightarrow \pi \eta\theta_{\rm eff}^M + 2\pi m$ 
in Eq. (\ref{Z_M}) and summing over $m \in \mathbb{Z}$. This is in fact a conventional procedure in similar systems, see for e.g.  
Eq. (52) in \cite{Thomas:2011ee} where theoretically controllable computations have been performed in a weakly coupled non-abelian gauge theory. 
 A generic consequence of this periodicity is that the system shows cusp singularities at specific  points, normally $\theta=\pi +2\pi m$. The origin for  such behavior is the   two-fold degeneracy emerging  in the system at these points. As a result of this degeneracy, level crossing occurs precisely at these points when the  branch with the minimum free energy (corresponding to the ground state)  changes , see \cite{Thomas:2011ee} for a large number of references on the original literature.}.  Using the periodicity, one can always separate out an integer multiple of $2\pi$ from stronger fields,  $ \eta \pi \theta_{\rm eff}^M =2\pi m +[ \eta \pi \theta_{\rm eff}^M-2\pi m]$, 
such that the pressure in the second branch, $ \pi\leq \eta \pi \theta_{\rm eff}^M \leq 2\pi$,
 assumes the form 
\be
 \label{theta_large}
 &&P_{\rm top} (\pi\leq \eta \pi \theta_{\rm eff}^M\leq 2\pi)
\approx \frac{e^2 \pi^2 \eta^2}{L_1^2 L_2^2} \exp{(-\pi^2\eta)}\\
&& \times\left[ \cos(\pi \eta \theta_{\rm eff}^M) +\frac{(2\pi -\pi\eta\theta_{\rm eff}^M)}{\pi^2\eta} \sin(2\pi -\pi\eta\theta_{\rm eff}^M) \right].  \nonumber   
\ee
Thus, the second term in the square brackets in (\ref{pressure_theta}) and (\ref{theta_large}) explicitly exhibits the cusp singularity at the point $\pi\eta\theta_{\rm eff}^M=\pi$. Such behavior is, in fact, a quite generic feature of gauge systems as a result of two-fold degeneracy and level-crossing phenomena,  see footnote \ref{degeneracy} for more comments and references. 

Indeed,  the degeneracy can be   observed from the partition function (\ref{Z_M}) which explicitly exhibits the double degeneracy  
at $\pi\eta\theta_{\rm eff}^M=\pi$  when the level crossing occurs. 
 One can approach the point $\pi\eta\theta_{\rm eff}^M=\pi$ from two sides using a mirror-like symmetry:
$\pi\eta\theta_{\rm eff}^M\rightarrow (2\pi-\pi\eta\theta_{\rm eff}^M)$.     \exclude{This symmetry can be checked by observing that  the   topological term in (\ref{Z_M}) flips the sign under the symmetry.} It can be easily checked that this does not modify  the partition function itself as 
the corresponding sign-flip in the topological term is equivalent to relabeling $k\rightarrow - k$ in the sum $k\in \mathbb{Z}$ in Eq. (\ref{Z_M}).
 An analogous phenomenon  is known to occur in the 2d Schwinger model where the system shows two-fold degeneracy at  $\theta =\pi$. It also occurs  in the 4d Maxwell system saturated by the magnetic instantons as  discussed   in details in \cite{Zhitnitsky:2013hba}. 
 
\begin{figure}
\centering
\includegraphics[width=0.47\textwidth]{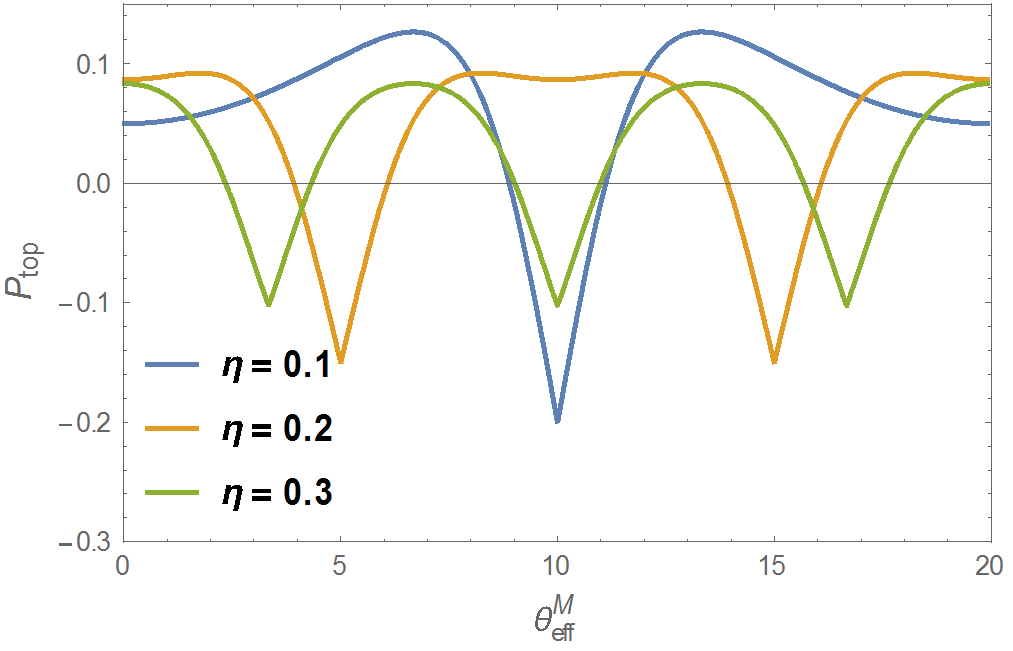}
\caption{\label{fig:pressure_theta} The topological Casimir pressure plotted in units of $e^2/2L_1^2 L_2^2$. There is a clear dependence of oscillation periodicity on $\eta$, in sharp contrast with the universal $2\pi$-periodicity in the topological pressure in the magnetic system \cite{Cao:2013na}. To make the pressure strictly periodic, Eq. (\ref{P_0}) is used in the first branch, $|\pi \eta\theta_{\rm eff}^M|<\pi$, and the results are then repeated over subsequent branches, see text with details.}
\end{figure}

 Numerically, for sufficiently small $\eta\sim 0.25$ where the effect is expected to be at its maximum,   the pressure  as a function of   $\theta_{\rm eff}^M$  is shown in
 Fig. \ref{fig:pressure_theta} and cusp singularities emerge at $\pi\eta\theta_{\rm eff}^M=\pi +2\pi m$. One can explicitly see that the system is symmetric under  $\pi\eta\theta_{\rm eff}^M\rightarrow 2\pi-\pi\eta\theta_{\rm eff}^M$, and it  also exhibits the periodicity  $\pi\eta\theta_{\rm eff}^M\rightarrow \pi\eta\theta_{\rm eff}^M +2\pi m$, as expected.

\exclude{ In fact, simply using Eq. (\ref{P_0}) to compute the induced pressure gives rise to unbounded behavior as $\theta_{\rm eff}^M$ increases. The reason is that $\theta_{\rm eff}^M \in \left\lbrace 1/\eta + 2n/\eta: n \in \mathbb{Z} \right\rbrace$ corresponds to the two-fold degeneracy points of the system where different energy levels cross and the ground state restructures itself \cite{Zhitnitsky:2013hba}. This unbounded behavior results from the fact that the differentiation in Eq. (\ref{P_0}) picks up the higher energy level after level crossing occurs, whereas all physically meaningful quantities must come from the lowest energy level. To restore the  $2/\eta$-periodicity that is apparent from the partition function (\ref{Z_M}), one could make the replacement 
\be
\label{periodicity}
\pi \eta\theta_{\rm eff}^M \rightarrow \pi \eta\theta_{\rm eff}^M + 2\pi m,  
\ee
where $m \in \mathbb{Z}$, and sum over $m$. Alternatively, we could restrict the use of Eq. (\ref{P_0}) to the first Brillouin zone only (before level crossing occurs), $|\pi \eta\theta_{\rm eff}^M|<\pi$, and repeat the results so obtained over subsequent Brillouin zones, as we have done in producing Fig. \ref{fig:pressure_theta}.}

\begin{figure}
\centering
\includegraphics[width=0.47\textwidth]{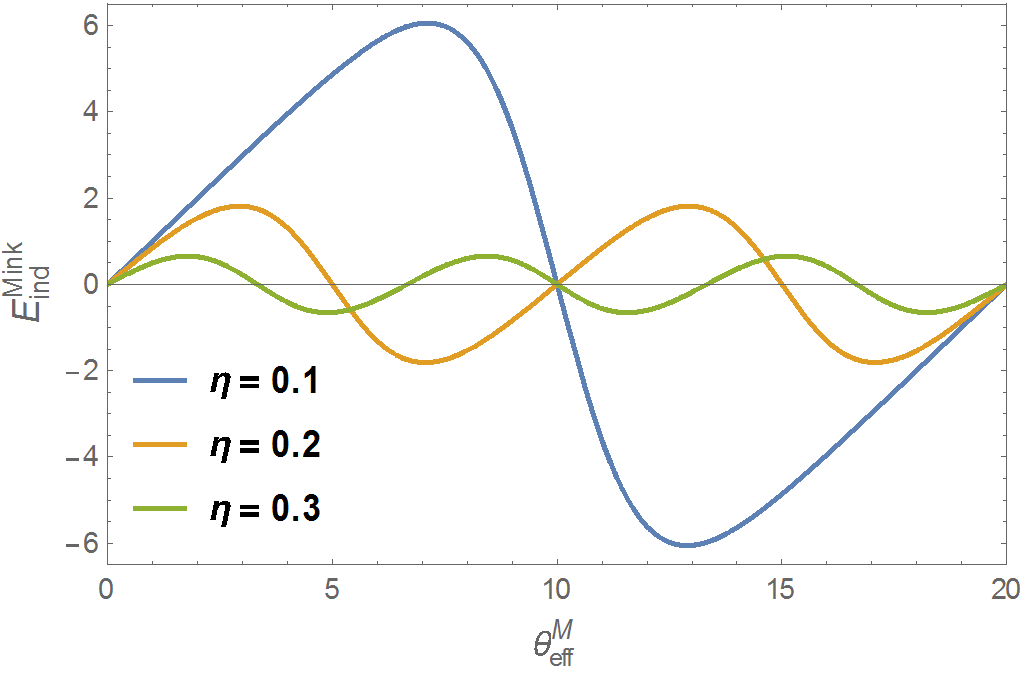}
\caption{\label{fig:E_ind} The induced electric field in units of $1/e \beta L_3$ in the presence of a external electric field in Minkowski space-time. The oscillation period and amplitude increase for smaller $\eta$. The same plot also serves as the induced dipole moment measured in units $\frac{L_1L_2L_3}{e \beta L_3}=\frac{e}{2}L_3\eta$, see sections \ref{dipole}, \ref{numerics} for the details. }
\end{figure}

One can also compute the induced electric field in response to the external source $\theta_{\rm eff}^M$ in Minkowski space by differentiating the partition function   $ \bar{{\cal{Z}}}_{\rm top}(\eta, \theta_{\rm eff}^M) $ with respect to $E_{\rm ext}^{\rm Mink}$:
\be 
\label{E_ind_M}
&\,& \langle E_{\rm ind}^{\rm Mink} \rangle = -\frac 1 {\beta V}\frac{\partial \ln \bar{\mathcal{Z}}_{\rm top}}{\partial E_{\rm ext}^{\rm Mink}}=
-\frac{e}{L_1 L_2}\frac{\partial \ln\bar{\mathcal{Z}}_{\rm top}}{\partial\theta_{\rm eff}^M}\\
&=& \frac{1}{\bar{\mathcal{Z}}_{\rm top}}\sum_{k\in\mathbb{Z}}\frac{2 \pi k}{ eL_3\beta} e^{-\eta\pi^2k^2}\sin \left[\pi k \eta \theta_{\rm eff}^M \right],\nonumber 
\ee
plotted in Fig. \ref{fig:E_ind}. 

It is quite obvious that $ \langle E_{\rm ind}^{\rm Mink} \rangle=0$ if $E_{\rm ext}^{\rm Mink}=0$. The difference in comparison with the analogous expression from \cite{Cao:2013na} is that our present definition of the induced field does not include the external piece $E_{\rm ext}^{\rm Mink}$, in contrast with the definition used in  \cite{Cao:2013na} where the induced field of the system was defined as the total field (i.e., the truly induced field plus the external piece). One can   observe that the induced field $ \langle E_{\rm ind}^{\rm Mink} \rangle$ changes the sign under the  mirror-like symmetry 
$\pi\eta\theta_{\rm eff}^M\rightarrow 2\pi-\pi\eta\theta_{\rm eff}^M$ in contrast with expression for the pressure presented  in Fig. \ref{fig:pressure_theta}. This is because the  induced field   $\sim {\partial}/{\partial\theta_{\rm eff}^M} $   flips the sign under  this  symmetry, in contrast with the pressure $P_{\rm top}\sim {\partial}/{\partial L_3}$.

A few comments are in order. First, formula (\ref{E_ind_M}) represents the physical induced field defined in Minkowski space-time. In addition, the oscillations now occur with periodicity  $\pi \eta\theta_{\rm eff}^M =2\pi m $,  which can be identically rewritten as 
\be
\label{E_quantization}
\pi \eta\theta_{\rm eff}^M =2\pi m ~~~~ \Rightarrow ~~~~ L_1L_2 E^{\rm Mink}_{\rm ext}=e m. 
\ee
  Note that this value for the physical electric field is indeed consistent with the results from the canonical Hamiltonian approach in Appendix \ref{Hamiltonian}. It further supports our formal manipulations in the transition from the Euclidean to Minkowski description. 
  
  \exclude{ The periodicity expressed by Eq. (\ref{E_quantization}) implies that points $\pi \eta\theta_{\rm eff}^M =\pi \pm 2\pi m$ are very special in a sense that 
   the system exhibits  an exact two-fold degeneracy at these specific values of the external filed, similar to the  magnetic case \cite{Zhitnitsky:2013hba}. This degeneracy in all respects is analogous to the well known degeneracy in two dimensional  Schwinger model at  $\theta=\pi$.   At this point the induced field flips the sign.  
}   
 We should emphasize that the quantization $L_1L_2 E^{\rm Mink}_{\rm ext}=e m$ of the physical electric field (\ref{E_quantization}) is drastically different from the quantization of the Euclidean instanton-fluxes (\ref{E_top}) saturating the partition function. Indeed, the instanton-fluxes have a different normalization factor  $e\rightarrow \frac{2\pi}{e}$ along with the geometric factor $L_1L_2\rightarrow L_3\beta$. It should be contrasted with the magnetic case  \cite{Cao:2013na} where the  external field and the instanton-fluxes  have  identically the same periodic properties.  This is precisely the reason why formulae from \cite{Cao:2013na}
 exhibit the universal periodic properties, in contrast with (\ref{E_ind_M}).     
 
In the large $\eta$ limit one can again keep only two terms in  Eq. (\ref{E_ind_M}) with $k=\pm 1$ such that it assumes the following simple form
      \be 
\label{E_eta}
  \langle E^{\rm Mink}_{\rm ind} \rangle = \frac{4 \pi e^{- \eta\pi^2 }}{ eL_3\beta}  \sin (\pi \eta \theta_{\rm eff}^M ). ~~
\ee 
As expected, the induced term has non-analytical $\exp (-1/e^2) $ behaviour and cannot be seen in perturbation theories as it originates from the tunneling events. It is exponentially suppressed when $\eta\rightarrow \infty$ as expected, and obviously vanishes in the $\theta_{\rm eff}^M\rightarrow 0$ limit.
 
\begin{figure}
\centering
\includegraphics[width=0.47\textwidth]{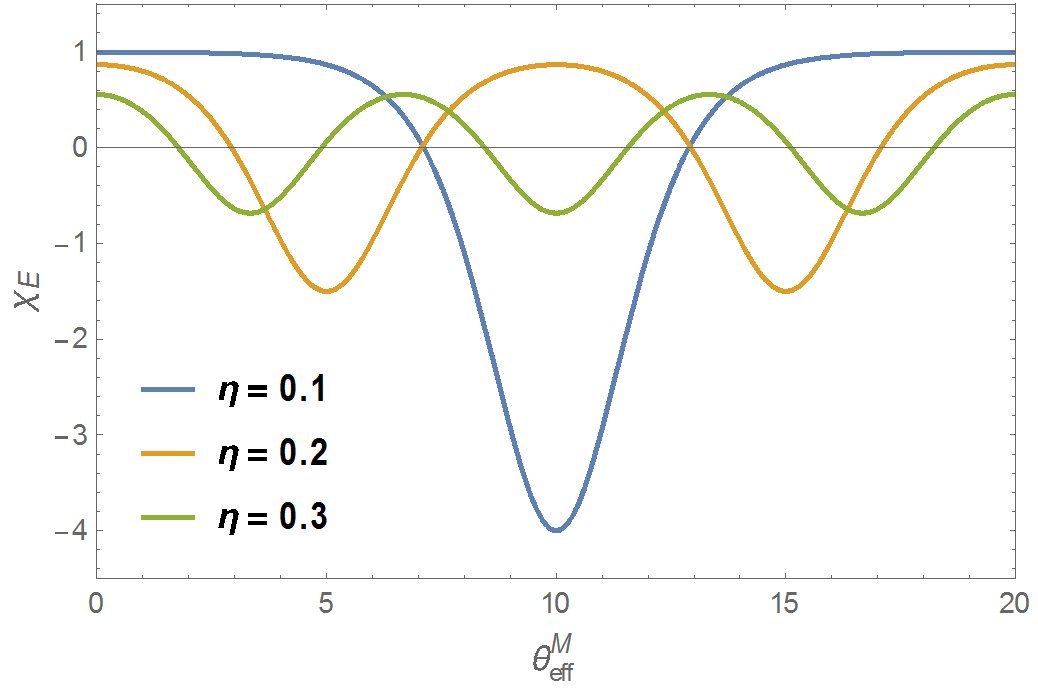}
\caption{\label{fig:susceptibility} The electric susceptibility of the system in response to different values of the external electric field. It is non-vanishing even at $\theta_{\text{eff}}^M=0$ due to topological fluctuations. The oscillations have $\eta$-dependent periodicity: $2/\eta$.}
\end{figure}

 Similarly, we can compute the (topological) susceptibility of the system, which measures the response of free energy to the introduction of a source term, represented by $E_{\rm{ext}}^{\rm Mink} \sim \theta_{\text{eff}}^M$ in our case:
\be
\chi_E =- \frac {2} {\eta} \frac{\partial^2}{\partial {\theta_{\text{eff}}^M}^2} \ln \bar{\cal{Z}}_{\text{top}}(\eta, \theta_{\text{eff}}^M),
\ee
shown in Fig. \ref{fig:susceptibility}.

In the large $\eta$ limit, one can explicitly check using the analytical expression (\ref{Z_M}) that $\chi_E \sim \exp[-\eta \pi^2]$ is strongly suppressed, consistent with our expectation that no electric correlations exist in the conventional Casimir effect setup.  The $\chi_E$ is symmetric under the  mirror-like symmetry 
$\pi\eta\theta_{\rm eff}^M\rightarrow 2\pi-\pi\eta\theta_{\rm eff}^M$ as  the second derivative ${\partial^2}/{\partial {\theta_{\text{eff}}^M}^2} $ 
does not flip  the sign  at the degeneracy point $\pi\eta\theta_{\rm eff}^M=\pi$. 

\section{Induced dipole moment and surface charges} \label{dipole}

Although the electric instanton-fluxes describing the tunneling events between topological sectors in our system are formulated in Euclidean space-time (\ref{E_top}), we have shown in the previous section that a real-valued electric field will be induced in Minkowski space-time in response to a non-vanishing $\theta_{\rm eff}^{\rm M}$. In this section, we represent this induced electric  field  in the bulk of the system  in  terms of the surface effects. In other words, we want to reformulate the fluctuations of the topological electric  fluxes using fluctuating surface charges on the plates. 

Similar reformulation of the problem was carried out for the magnetic case in \cite{Zhitnitsky:2015fpa} where it was  explicitly shown that the tunneling   instanton effects can be understood in terms of fluctuating topological non-dissipating   currents which unavoidably will be generated on the boundaries. In  \cite{Zhitnitsky:2015fpa},  the system 
was  a   cylinder with radius $R$  and height  $L_3$  with  the  topological current $J^{\phi}$  flowing  in the $\phi$-direction  on its infinitely thin boundary. It has also been noted  that  these topological currents are very similar in nature, but not the same as the well-known  persistent currents normally observed on metallic (not superconducting) rings, see references on the original experimental and theoretical studies on persistent currents in  \cite{Zhitnitsky:2015fpa}.  In all cases  the effects are  due to the coherent Aharonov-Bohm phases  correlated on macroscopically large distances, although the nature 
of the long-range coherence is different  for our topological non-dissipating currents and for the well-known persistent currents.

The duality between magnetic and electric fields in the Maxwell system strongly suggests that similar topological effects must be present in the electric systems as well.  Essentially we attempt to study the effects which are EM-dual 
to the persistent currents observed on metallic rings. The relevant formal construction indeed can be easily carried out as we have shown in sections \ref{electric} and \ref{external field}.

We start with an explicit  demonstration  that 
the existence of an induced electric field (generated by the electric instantons)  in Minkowski space (\ref{E_ind_M})  suggests that one can effectively recast the mathematics into an equivalent form where physical electric  charges
are induced on each plate of the capacitor\footnote{We briefly remark that the charges indeed physically reside on the actual plates as long as the periodic boundary conditions are enforced using the quantum LC circuit configuration.}. These charges have pure quantum origin and are, in all respects, very similar to the persistent non-dissipating currents induced by the magnetic instantons discussed in \cite{Zhitnitsky:2015fpa}. 

An important technical comment here is that the induced electric  field  (\ref{E_ind_M})  can be thought of as the polarization of the system per unit volume, i.e. $\la P\ra=-\la E^{\rm Mink}_{\rm ind}\ra$, since the definition for 
$\la P\ra$ is identical to (\ref{E_ind_M}) up to a minus sign because it enters the Hamiltonian as $H=-\vec{P} \cdot \vec{E}_{\rm ext}$. Therefore, we arrive at the following expression for the induced electric dipole moment of the system  in the presence of an external electric field $E_{\rm ext}^z$, 
 \be
\label{p_ind_M}
&\,& \langle p^{\rm Mink}_{\rm ind} \rangle = -\langle E^{\rm Mink}_{\rm ind} \rangle   L_1L_2L_3 \\ 
&=&  -\frac{1}{ \bar{{\cal{Z}}}_{\rm top}} \sum_{k\in\mathbb{Z}}\frac{2 \pi k L_1 L_2}{ e\beta} e^{-\eta\pi^2k^2}\sin (\pi k \eta \theta_{\rm eff}^M ). \nonumber
  \ee
  Based on this interpretation,  one can view Fig. \ref{fig:E_ind} as a plot for the induced electric dipole moment in units of $
  \frac{1}{2}\eta eL_3$
  which represents the correct dimensionality  ${\rm e}\cdot {\rm cm}$ for the electric dipole moment. 
  
One can understand the same formula (\ref{p_ind_M}) using the original expression for the coupling between the external field and the topological instantons
  \be
\label{E_cross}
 \int \dd^4 x  \left(\vec{E}_{\rm ext} \cdot \vec{E}_{\rm top}\right)=   \int \dd^4 x ~{V}_{\rm ext} \left(\vec{\nabla}\cdot \vec{E}_{\rm top}\right),~~
\ee
  where we have neglected a  total  divergence term. The cross term written in the form (\ref{E_cross}) strongly suggests that 
   $  \left(\vec{\nabla}\cdot \vec{E}_{\rm top}\right)$ can be interpreted as surface charges generated on the plates. Indeed, if we use
   Eq. (\ref{E_top}) for the topological instantons  $ \vec{E}_{\rm top}$ describing the tunneling transition to the $|k\ra$ sector, we arrive at the following formula for the surface charge density 
  \be
  \label{charge}
\sigma^{\rm Mink}_{\rm ind}(k)=\frac{2 \pi k}{e\beta L_3}\left[ \delta (x_3)-\delta(x_3-L_3) \right].      
\ee
  This formula implies that an electric dipole moment will be generated  in each topological sector $|k\ra$, given by 
 \be
 \label{dipole_k}
p^{\rm Mink}_{\rm ind}(k)=-\frac{2 \pi L_1L_2}{e\beta}k, 
  \ee
 which reproduces the relevant term in (\ref{p_ind_M}), derived in a quite different way without any mention of surface charges. 
 
 Furthermore, one can explicitly check that the cross term (\ref{E_cross}) expressed in terms of the surface charge $   \sigma^{\rm Mink}_{\rm ind}(k) $ exactly reproduces the corresponding term in the action of the partition function (\ref{Z_M}) computed in terms of the bulk instantons. Indeed, the cross term in the action is 
   \be
\label{E_cross1}
 \int \dd^4 x  \left(\vec{E}_{\rm ext} \cdot \vec{E}_{\rm top}\right)=   \int \dd^4 x ~{V}_{\rm ext} \left(\vec{\nabla}\cdot \vec{E}_{\rm top}\right)~~\\
 = \int \dd^4 x  \left({E}_{\rm ext} x_3\right) \left(\frac{2 \pi k}{e\beta L_3}\right) \delta(x_3-L_3) =\pi \eta k \theta_{\rm eff}^E , \nonumber
\ee
where we have substituted ${V}_{\rm ext}=-x_3 {E}_{\rm ext}$ and expressed the external field in terms of $\theta_{\rm eff}^E$.
 Eq. (\ref{E_cross1})  precisely coincides with the cross term $\sim k$ in the action $\exp(-S)$    for the partition function (\ref{normalization}). In the same way the classical instanton action $\sim k^2$ in (\ref{normalization}) can be also understood in terms of the surface charges. 
 
 The basic point of our discussions in this section is that the expression for the induced electric dipole  moment (\ref{p_ind_M}) can be understood  in terms of an induced field  according to (\ref{E_ind_M}). The same effect can be also interpreted in terms of the surface charges  of the system as (\ref{charge}) and (\ref{dipole_k}) suggest. However, 
 the origin  of the phenomena is not these charges  but  the presence of the 
  topological $|k\ra$ sectors in Maxwell $U(1)$ electrodynamics formulated on a compact manifold with nontrivial  
  mappings $\pi_1 [U(1)]=\mathbb{Z}$.  Such $|k\ra$ sectors exist and transitions between them always occur  even if  charged   particles are not present in the bulk of the system.  The secondary role played by the charged particles is in particular illustrated by the fact that the extra contribution to the Casimir vacuum pressure generated by $\bar{{\cal{Z}}}_{\rm top}$ survives a vanishing external field, but the fluctuating positive and negative charges cancel each other exactly.
  The fundamental explanation is still the tunneling transitions between vacuum winding states which occur regardless of the value of the external field.   
  
  To conclude this section, we would like to remark that it is quite typical in condensed matter physics that topologically ordered systems allow such a complementary formulation in terms of the physics on the boundary. Therefore, it is not a surprise that we can reformulate the original instanton fluctuations saturating $  \bar{{\cal{Z}}}_{\rm top}(\eta, \theta_{\rm eff}^E) $ in terms of  the boundary surface charges which always accompany these instanton transitions. See more discussions on the relation between descriptions in the bulk and on the surface
  for the magnetic system  in \cite{Zhitnitsky:2015fpa}.

\section{Numerical estimates for a quantum LC circuit}\label{numerics}
Our goal now is to discuss a possible  design where such electric  effects can be  (at least in principle) studied. In what follows we consider a two-plate capacitor with area $L_1L_2$ separated by distance $L_3 \ll L_1, L_2$ at temperature $\beta^{-1}$. The two plates are connected by an external wire such that charges can freely move  from one plate to the other.  The system can be viewed as a quantum LC circuit when conventional quantum transitions (due to ordinary  degrees of freedom)  are replaced by tunneling transitions as described in previous sections. The vacuum  between the plates (where these tunneling transitions occur) represents  the  object of our studies. 

We would like to make a few numerical estimates for illustration purposes only.
The first set of parameters is motivated by the  accurate measurement  of the CE using parallel plates \cite{Casimir-exp} (see also 
\cite{Casimir-exp1} where historically the first accurate measurement was performed but for a different geometry).
The second set of parameters is motivated by the experiments  on persistent currents \cite{persistent-exp} where the correlation of the Aharonov-Bohm phases is known to be maintained. While the persistent current is a magnetic phenomenon, the EM duality strongly suggests that a similar electric effect should also occur when the coherent Aharonov-Bohm phases are correlated on macroscopically large distances. Therefore, for the second set of parameters we adopt the typical sizes of the magnetic system (where persistent currents have been observed) to estimate the topological effects for our electric capacitor.
  
First, we adopt the system  sizes  from the first accurate measurement of the Casimir effect using parallel plates \cite{Casimir-exp}. 
Unlike the magnetic system studied in \cite{Cao:2013na}, 
the dimensions of this electric system as represented by $\eta$ can be fairly easily optimized to maximize the topological effect. 
In this experiment a small capacitor was formed using chromium-coated surfaces with area $L_1L_2=1.2 \times \SI{1.2}{\milli\meter\squared}$ and separation $L_3$ in the $0.5-\SI{3}{\micro\meter}$ range. This geometry corresponds to $\eta \gg 1$ and for reasons explained previously, we expect the topological effect to be vanishingly small. However, if plate separation could be increased to \SI{0.4}{\milli\meter} (such that our approximation $L_3 \ll L_1, L_2$ is still marginally satisfied) 
  and the ambient temperature set to \SI{10}{\milli\kelvin} (corresponding to $\beta = \SI{180}{\milli\meter}$), then 
\be
\label{eta1}
\eta^{(I)}=\frac{2 L_1 L_2}{e^2 \beta L_3}=\frac{1.2 \times \SI{1.2}{\milli\meter\squared}}{2\pi\alpha (\SI{180}{\milli\meter}) (\SI{0.4}{\milli\meter})}\approx0.4,
\ee
where the intensity of the topological tunneling transitions assumes the  maximum   values, see Fig. \ref{fig:pressure_0}.
The effect is still much smaller than conventional Casimir effect as we already discussed in section \ref{pressure}. 
The main point, however, is that the effect is highly sensitive to the external field, in contrast with the conventional CE,  as argued in section \ref{external field}. Precisely this sensitivity 
might be the key element for observing this fundamentally novel phenomenon when the vacuum energy in the bulk is not associated   with any physical  degrees of freedom propagating in the bulk.

Now we want to consider a second set of parameters motivated by observing the persistent currents \cite{persistent-exp}.
 It has been mentioned previously \cite{Zhitnitsky:2015fpa} that the topological non-dissipating currents, while  similar in nature to the observed persistent currents  \cite{persistent-exp}, are nevertheless not identically the same and represent an independent additional contribution to the non-dissipating currents\footnote{In the conventional case,
 the Aharonov-Bohm coherence is determined by the dynamics of the electrons residing on the ring, while in our case it is determined by the dynamics of the vacuum, i.e. tunneling transitions between the  $|k\ra$ winding  sectors.  This difference, in particular, manifests itself  in the properties of the induced magnetic moment: it is quantized in our case, while it can assume any value for conventional persistent currents.}. Based on EM duality, one could argue  that if Aharonov-Bohm coherence has been established for  the magnetic system, it is likely to hold in electric systems with similar geometric sizes as well.  Therefore, we expect the topological tunneling transitions to be present in the system when the ring area $\pi R^2$ in the magnetic system \cite{persistent-exp} is replaced by the  electric capacitor plate area $L_1L_2$, and the ring width replaced by the plate separation $L_3$. With this correspondence,  we estimate the key dimensionless parameter $\eta$ to be
 \be
\label{eta2}
\eta^{(II)}=\frac{2 L_1 L_2}{e^2 \beta L_3}=\frac{2\pi (1.2{\mu m})^2}{4\pi\alpha (0.6 \rm cm) (0.1 \mu m )}\approx0.16,
\ee
where we have used $L_3\sim 0.1 \mu {\rm m} $ and $\beta\sim 0.6$ cm corresponding to the temperature $T\simeq 300 $ mK
below which the electron phase coherence length  is sufficiently large and temperature-independent\footnote{\label{AB}One should remark here that there are related effects  when  the entire system  can   maintain coherent Aharonov-Bohm phases at very  high temperatures $T\simeq  79 $ K \cite{persistent-temp}.}. This parameter  falls into the region 
where the intensity of the topological tunneling transitions assumes its  maximum   values, see Fig. \ref{fig:pressure_0}.
Therefore, one should anticipate a nonzero value for the induced electric dipole moment when an external field is applied.

Assuming appropriate boundary conditions (i.e., periodic up to a large gauge transformation) are established, the induced electric dipole moment depends on the applied external field. The corresponding dependence of $\langle p^{\rm Mink}_{\rm ind} \rangle$ on external field can be easily established  from Fig. \ref{fig:E_ind} where the plot should be understood as  the induced dipole moment in units of $E_{\rm ind}L_1L_2L_3= \frac{eL_3}{2}\eta$. Numerically, these units   for our  two sets of parameters  can be estimated as 
 \be
 \label{dipole_numerics}
\langle p^{\rm Mink}_{\rm ind} \rangle^{(I)} &\approx& \frac{eL_3}{2}\eta^{(I)}\sim 0.1 (e \cdot{\rm mm})\nonumber \\
\langle p^{\rm Mink}_{\rm ind} \rangle^{(II)} &\approx& \frac{eL_3}{2}\eta^{(II)}\sim 0.01 (e \cdot{\rm \mu m}).
\ee
The numerical estimates  for the induced dipole moment due to the coherent tunneling effects (\ref{dipole_numerics}) do not look very promising if measurements are performed with a static external electric field. In the next section we consider an option when the external field varies with time. In this case one should anticipate the emission of real propagating photons which leave the system, and which hopefully (in principle) can be detected.  

\section{Electrodynamics and E\&M radiation}\label{radiation}

Up till this point, we have implicitly assumed that the external electric field $E_{\rm ext}^{\rm Mink}$ is static. However, Eq. (\ref{p_ind_M}) still holds in the case of a dynamical external field $E_{\rm ext}^{\rm Mink} (t)$ as long as its time dependence is adiabatically slow compared to all relevant time scales of the system. Then this time-dependence simply enters the partition function through $\theta_{\rm eff}^M(t)$. In has been argued in \cite{Zhitnitsky:2015fpa} that as the induced magnetic dipole moment varies in response to an adiabatically oscillating external source, real photons will be emitted through magnetic dipole radiation. Following the same reasoning, we can also compute the electric dipole radiation produced by our capacitor. 

  Therefore, one can use the well-known expressions for the intensity $  \vec{S}$ and total radiated power $I$ for the  electric dipole radiation when the dipole moment (\ref{p_ind_M}) varies with time:
\be
\label{intensity}
  \vec{S} = I(t) \frac{\sin^2 \theta}{4\pi r^2}\vec{n}, ~~~~~~~~~   I (t) =\frac{2}{3 c^2} \langle \ddot{p}^{\rm Mink}_{\rm ind} (t)\rangle^2
\ee
In the case where the external electric field plotted in Fig. \ref{fig:E_ind}  is almost linear, the approximate induced dipole moment,  $E_{\rm ext}^z(t) $, varies linearly  as one can see from Fig.\ref{fig:E_ind}. In particular,  if $E_{\rm ext}^z(t)\sim \cos\omega t$
then $\langle \ddot{p}^{\rm Mink}_{\rm ind} \rangle \sim \omega^2\cos \omega t$. In this case one can easily compute the average intensity over large number of completed cycles  with the result 
\be
\label{intensity1}
  \la I \ra \sim \frac{\omega^4}{3 c^2} \langle {p}^{\rm Mink}_{\rm ind} \rangle^2,
\ee
where $ \langle {p}^{\rm Mink}_{\rm ind} \rangle$ is given by (\ref{p_ind_M}).

\exclude{
In particular, for a sinusoidally oscillating external field $E_{\rm ext}^{\rm Mink} (t) \sim \cos(\omega t)$, we have $\langle \ddot{p}_{\rm ind}^{\rm Mink} \rangle \sim \omega^2 \cos(\omega t)$, where $\langle p_{\rm ind}^{\rm Mink} \rangle$ is given by (\ref{p_ind_M}). Then, the radiation intensity is
\be
\vec{S} (t)= \frac{\mu_0}{16 \pi^2 c} \frac{\sin^2\theta}{r^2} \langle \ddot{p}_{\rm ind}^{\rm Mink} \rangle^2 \vec{n} \sim \frac{\sin^2\theta}{r^2} \cos(\omega t) \vec{n}.
\ee
}
This electromagnetic radiation has a simple explanation in terms of the fluctuating charges that reside on the capacitor plates (\ref{charge}). For static external field $E_{\rm ext}^{\rm Mink}$, the charge density on each plate also stays constant, giving rise to a time-independent electric dipole moment. But when $E_{\rm ext}^{\rm Mink} (t)$ starts to fluctuate, the induced dipole moment also acquires time-dependence, naturally leading to the emission of real photons. 
\exclude{It is worth mentioning that, as in the magnetic case, this explanation in terms of fluctuating charges or currents is the consequential, rather than fundamental explanation for the emission. The fundamental reason is still the instanton-fluxes interpolating between topological sector $|k\ra$, which exist regardless of any external source.
}

The detection of this radiation could be made easier by taking advantage of the system's two-fold degeneracy at $\theta_{\rm eff}^M \in \left\lbrace 1/\eta + 2n/\eta: n \in \mathbb{Z} \right\rbrace$ corresponding to half integer fluxes, where the system's ground state reconstructs itself \cite{Zhitnitsky:2013hba}. As the external field slowly sweeps through these points, $\langle p_{\rm ind}^{\rm Mink} \rangle$ quickly changes polarity as Fig. \ref{fig:E_ind} shows, producing high amounts of radiation. This will hopefully set the topological emission apart from the uninteresting radiation due to the fluctuating external source. In addition, since the power radiated electrically usually substantially exceeds that radiated magnetically for systems with comparable dimensions, we expect the radiation from this present electric configuration to be much more readily detectable than the one in \cite{Zhitnitsky:2015fpa}.

 \section{Conclusion and Speculations }\label{conclusion}
In this work we have discussed a number of very unusual features   exhibited by    the   Maxwell theory formulated on  a compact manifold $\mathbb{M}$ with nontrivial topological mappings $\pi_1[U(1)]$, which was termed the  topological vacuum ($\cal{TV}$).  All these  features originate from the topological portion of the partition function ${\cal{Z}}_{\rm top}$
 which cannot be described by a construction of conventional photons expanded near a classical vacuum $|n\rangle$.  In other words, all effects discussed in this paper have a non-dispersive nature. 
  
             The computations of the present work along with previous calculations in \cite{Cao:2013na,Zhitnitsky:2013hba,Zhitnitsky:2014dra,Zhitnitsky:2015fpa}  imply  that the extra energy,   not associated   with any physical propagating degrees of freedom,  may emerge  in a  gauge  system if some conditions are met. This fundamentally new type  of  energy    emerges as a result of the dynamics of pure gauge configurations. 
             
             The new idea advocated in this work is that this new type of energy 
             might be related to  electric-type instantons, in contrast with magnetic-type instantons studied in the  previous papers.
             While the modification may look minor, it leads to a number of novel effects. 
             Most notably, whereas the magnetic oscillations have universal $2\pi$-periodicity for all system sizes, the periodicity in the electric system varies with the system size. 
 It remains  to be seen if the effects discussed in the present work can be experimentally observed. 
 
In addition, 
     the electric system studied in the present work could be considered as  a  simple toy  model where the topological tunneling transitions can be theoretically studied. These novel effects  could have  some profound consequences for  astrophysics  and cosmology if they persist in the Standard Model (SM).
     Indeed, the concept of gauge symmetry is central to the SM. As a result, the existence of such non-trivial homotopy mappings implies that the ground state of the system should be represented by a superposition of the topological winding sectors. The corresponding path integral which determines the system therefore must also include the sum over all the topological sectors, accounting for the physics describing the tunneling transitions between them. 
     
     All   the  effects studied in sections \ref{electric} and \ref{external field}  were precisely attributed to the   physics of tunneling transitions which generate an additional vacuum energy. The effects are obviously non-local in nature. Moreover, the corresponding contributions to the correlation functions  are  represented by non-dispersive terms which fundamentally cannot be expressed in terms of any local propagating degrees of freedom. Furthermore, the radiation of real physical photons described in section \ref{radiation}
     represents a fundamentally novel mechanism of particle production (\`{a} la the dynamical Casimir effect) when the emission occurs from the topological vacuum due to the tunneling transitions, rather than from the decay  of some local   propagating degrees of freedom.
     
     We note that many conventional approaches to describe the dark vacuum energy (or vacuum inflation) are based on effective local field theories.   
     The production of real particles (during the so-called reheating epoch)  is also normally considered as a result of conventional local interaction of real particles.

 The unique features of the system studied in the present work when an extra energy is not related to any physical propagating degrees of freedom was the main  motivation for a proposal   \cite{Zhitnitsky:2013pna,Zhitnitsky:2015dia}  that the  vacuum energy of the Universe may have, in fact,  precisely such non-dispersive and non-local  nature\footnote{This new type  of vacuum energy which cannot be expressed in terms of propagating degrees of freedom is well known to the QCD community and has been   well studied in QCD lattice simulations, see  \cite{Zhitnitsky:2013pna,Zhitnitsky:2015dia} for references and details.} studied in the present work in a toy model. This proposal where an extra energy   cannot be associated with any propagating particles  should be contrasted with the conventional description where an extra vacuum energy in the Universe is always associated with some  ad hoc   physical propagating degrees of freedom, such as inflaton\footnote{There are two instances in the evolution of the universe when the vacuum energy plays a crucial  role.
 The first instance is identified with  the inflationary epoch  when the Hubble constant $H$ was almost constant, which corresponds to the de Sitter-type behavior $a(t)\sim \exp(Ht)$ with exponential growth of the size $a(t)$ of the Universe. The  second instance when the vacuum energy plays a dominant role  corresponds to the present epoch when the vacuum energy is identified with the so-called dark energy $\rho_{DE}$
 which constitutes almost $70\%$ of the critical density. In the proposal  \cite{Zhitnitsky:2013pna,Zhitnitsky:2015dia}  the vacuum energy density can be estimated as $\rho_{DE}\sim H\Lambda^3_{QCD}\sim (10^{-4}{\rm  eV})^4$, which is amazingly  close to the observed value. }. 
 \section*{Acknowledgements} 
     This research was supported in part by the Natural Sciences and Engineering Research Council of Canada. C.C. was in part supported by the Walter Burke Institute for Theoretical Physics at Caltech, the DOE grant DE-SC0011632, and the Gordon and Betty Moore Foundation through Grant 776 to the Caltech Moore Center for Theoretical Cosmology and Physics. Y.Y. acknowledges a research award from the UBC Work Learn Program.

\appendix
\section{Hamiltonian Approach}\label{Hamiltonian}
We check the topological partition function with external electric field using canonical quantization so there is no ambiguity regarding the physical meaning of the electric field.
Recall the classical lagrangian density for the EM field 
\begin{equation}
 \mathcal{L}(x)=\frac 1 2( \textbf{E}^2 -\textbf{B}^2),
\end{equation}
where $E_i=-\partial_i A_0 + \partial_0 A_i$ and $B_i=-\epsilon_{ijk}\partial_j A_k$.
We then proceed with the usual Hamiltonian formalism where we realize that $A_0$ is non-dynamical and imposes the Gauss' law constraint as a Lagrange multiplier. Hence the Hamiltonian density is given by
\begin{equation}
 \mathcal{H}(x)=\frac 1 2 (\textbf{E}^2 +\textbf{B}^2)- A_0(x)\nabla\cdot \textbf{E}(x).
\end{equation}
We can fix the $A_0=0$ gauge and define the Hamiltonian $H=\int d^3x \mathcal{H}(x)$.

Then, we follow the usual quantization procedure for the EM field by imposing the equal time commutation relations

\begin{equation}
\label{commutator}
 [\hat{A}_j(\textbf{x}),\hat{\Pi}_k(\textbf{y})] = i\delta(\textbf{x}-\textbf{y})\delta_{jk}.
\end{equation}

The conjugate momentum $\hat{\Pi}_i(x)=-\hat{E}_i$ and potential $\hat{A}_{i}$ are now promoted to operators, and 
\begin{equation}
 \hat{E}_{j}=i \frac{\delta}{\delta A_{j}(x)}.
\end{equation}

We also note the potentials related by a local time-independent gauge transformation are physically equivalent. Namely, $[\hat{Q}(\textbf{x}), \hat{H}]=0$, where $\hat{Q}(\textbf{x})=\nabla \cdot \textbf{E}(\textbf{x})$ generates local time-independent gauge transformations.
Consequently, we restrict physical states to the set of gauge invariant subspace of the original Hilbert space and $\hat{Q}(\textbf{x})|\text{phys}\rangle=0$.
Note that so far we haven't imposed any non-trivial boundary conditions.

Now we impose the non-trivial $S^1$ topology by requiring periodicity in the $L_3$ direction. Then, in addition to the original small gauge redundancy, we now also have large gauge transformations in the form of 
\be
\label{A_3}
A_3 \rightarrow A_3+2\pi n/eL_3.
\ee 
Unfortunately, the requirement on only small gauge transformations $[\hat{Q}(x), \hat{H}]=0$ is no longer sufficient for overall gauge fixing. 

This means that the ``physical states'' from the previous construction with trivial topology now have to be grouped into sectors. We put quotes because they are not really physical anymore as in the previous case. So instead of using a universal label $|\text{phys}\rangle$ for the set of ``physical states'', let us label each set according to their respective sectors, namely $|\text{phys}_n\rangle$ for the $n$-th sector. In particular, we can see that for each sector, such ``physical states''  indeed satisfy $\hat{Q(\textbf{x})}|\text{phys}_n\rangle =0$, but under large gauge transformation, we have $\mathcal{T}^{(k)}|\text{phys}_n\rangle=|\text{phys}_{n+k}\rangle$. The large gauge transformation operator $\mathcal{T}^{(k)}$ commutes with the Hamiltonian $[\mathcal{T}^{(k)}, \hat{H}]=0$. Therefore any physical observables computed without taking into account the sectors are not strictly correct. Furthermore, one has to be careful in not fixing this gauge but sum over all the inequivalent topological $n$ sectors, as they do give rise to physically measurable effects. 

As such, one needs to further modify our definition of the physical states (and/or associated Hilbert space). For clarity, we denote the ``vacuum'' state of each sector by $|n\rangle$ and define $|\theta\rangle = \sum \exp(i\theta n)|n\rangle$ as the new set of physical states which are indeed invariant under all large gauge transformations. The vacuum expectation for a physical observable is then given by$\langle \theta ' |\hat{A}|\theta\rangle = \delta(\theta '-\theta) \langle \theta |\hat{A}|\theta\rangle$. It is clear that different values of $\theta$ now label the superselection sectors of the theory.

Now we can compute the correct vacuum to vacuum transition (dropping the delta function) from $t=0$ to $t=t_f$. The Hamiltonian is time-independent, thus we get $\langle\theta| e^{-i\hat{H}t_f} |\theta\rangle$. This, as we all know, is the same as performing a related Feynman path integral. The Hamiltonian operator is given by 

\begin{equation}
\hat{H} = \int d^3 x \frac 1 2 (-\sum_i\frac{\delta^2}{\delta A_i^2} +\textbf{B}^2)
\end{equation}
%where, as usual, $B_i = \epsilon_{ijk}\partial_j \hat{A}_k$.
and the partition function is therefore
\begin{equation}
 {\cal Z}=\Tr(e^{-\beta \hat{H}})
\end{equation}
In 2d it is clear how to proceed as one can solve for the wavefunctional of $A$. The 4d computation requires more work due to the physically propagating degrees of freedom.

 Nevertheless, we can check the calculation for external electric fields. Adding a physical electric field in the hamiltonian is straightforward by letting, 
\begin{equation}
 \hat{E}_3=i\frac{\delta}{\delta A_3} \rightarrow i\frac{\delta}{\delta A_3} + E^{\rm M}_{\text{ext}},
\end{equation}
where $E^{\rm M}_{\text{ext}}$ is a physical classical electric field (times identity operator) in the $z$-direction. If we diagonalize the matrix according to its appropriate energy eigenfunction(al)s, then
\begin{equation}
 {\cal Z}=\exp(-\frac 1 2 \beta V (E^{\rm M}_{\text{ext}})^2) \times \Tr\exp(-\beta\hat{H}_{\text{sys}}) ={\cal Z}_{\text{ext}}\times {\cal Z}_{\text{sys}}
\end{equation}
where the first factor is nothing but the Boltzmann factor due to the external(applied) electric field in the system, and the Hamiltonian operator inside the trace now assumes the form
\begin{equation}
\hat{H}_{\text{sys}} = \int d^3 x \frac 1 2 (-\sum_i\frac{\delta^2}{\delta A_i^2} +\textbf{B}^2 + 2iE^{\rm M}_{\text{ext}}\frac{\delta}{\delta A_3}+\dots).
\label{eqn:modham}
\end{equation}
Note that we can factor out this external part, because it doesn't depend on the quantum fluctuations nor the topological sectors. As ${\cal Z}_{\text{ext}}$ is not relevant to our discussion, it suffices to preserve the second factor, ${\cal Z}_{\text{sys}}$, for analysis of the system.

In particular, we recognize the usual trick and evaluate the quantum partition function using an Euclidean path integral. We obtain the Lagrangian in the usual way and then perform a Wick rotation.
\begin{align}
{\cal Z}_{sys} &={\cal Z}_{\text{top}}\times {\cal Z}_{\text{quant}}\\ &=\sum_{k\in\mathbb{Z}}\int \mathcal{D}A^{(k)} \exp(-\int_0^{\beta} d\tau \int d^3x \mathcal{L}_E[A^{(k)}])
\end{align}
where $A^{(k)}$ is understood to be periodic in $\beta$, $k$ labels the topological sector, and the Lorentz indices are suppressed. The summation as well as the path integral over $A^{(k)}$ follow from the trace operation. Physical quantities such as $\langle E_{\text{ind}}\rangle$ and $\chi_E$ can thus be calculated by differentiating with respect to the source $E^{\rm M}_{\text{ext}}$. As demonstrated in section \ref{electric}, we expect the topological contribution to decouple from the conventional propagating photons.

 In particular, we treat $E^{\rm M}_{\text{ext}}$ as a constant classical field (or fixed parameter) that is coupled with $E_3$. The (cross) term in the (Minkowski) action that will contribute non-trivially to the final topological action takes the following form
\begin{equation}
iS_c^{\rm M}=i\int dt \int d^3x E_3^{\rm M} E^{\rm M}_{\text{ext}}.
\end{equation}

Performing a Wick rotation, we get
\begin{equation}
i\int d\tau \int d^3x E_3^{\rm E} E^{\rm M}_{\text{ext}}.
\end{equation}

One can also analytically continue $E^{\rm M}_{\text{ext}}$ in the path-integral for the actual computation. Because the topological contribution ultimately decouples from the functional integration, we have 

\begin{align}
{\cal Z}_{top} &= \sum_{k\in\mathbb{Z}} \exp(-\pi^2\eta k^2+i\frac{2\pi k E^{\rm M}_{\text{ext}}}{e\beta L_3}V\beta)\\
 &= \sum_{k\in\mathbb{Z}} \exp(-\pi^2\eta k^2+i\eta\pi k\theta_{\text{eff}}^M).
\end{align}
which is indeed in agreement with (\ref{Z_M}). Then (\ref{E_ind_M}), (\ref{E_quantization}) automatically follow. Note that the partition function is invariant under $\theta^M_{\text{ext}} \rightarrow -\theta^M_{\text{ext}}$ (which corresponds to $k\rightarrow -k$).

Another way to understand the same feature of the quantization (\ref{E_quantization}) of the electric field in the Hamiltonian approach is as follows. The large gauge transformations (\ref{A_3}) imply that the combination $eA_3L_3$ must be treated as a phase $\phi$ which is an angular variable. At the same time the commutation relation (\ref{commutator}) after integrating over $\int d^3xd^3y$ 
can be rewritten as follows
\be
\label{commutator1}
 \left[\frac{e\hat{A}_3 V}{L_1L_2},\tilde{E}_3 V \right] =i\frac{e}{L_1L_2}V
\ee
where $V$ is the volume of the system $V=L_1L_2L_3$ and $\tilde{E}_3$ is the operator of the electric field along the $z$-direction.
Commutator (\ref{commutator1}) can be written in the canonical form
\be
\label{commutator2}
 \left[\phi,\tilde{E}_3 \right] =i\frac{e}{L_1L_2}, ~~~~ \phi\equiv eA_3L_3,
\ee
 which implies that the operator for the electric field can be represented as follows
 \be
 \label{E_3}
 \tilde{E}_3=-i\left(\frac{e}{L_1L_2}\right)\frac{\partial}{\partial \phi}, 
 \ee
 similar to angular momentum operator $l_3$. The corresponding eigenfunctions have  the form $\sim \exp(im\phi)$, while
 the eigenvalues for the electric field are  
 \be
 \label{quantization_hamiltonian}
L_1L_2 \tilde{E}_3=em, 
 \ee
 which precisely coincides with quantization (\ref{E_quantization}).

\end{document}